\documentclass[conference]{IEEEtran}  

\usepackage{ifpdf}
\usepackage{cite}
\ifCLASSINFOpdf
   \usepackage[pdftex]{graphicx}
\else
    \usepackage[dvips]{graphicx}
\fi
\usepackage[cmex10]{amsmath}
\usepackage{amssymb}
\usepackage{amsthm}
\usepackage{graphicx}
\usepackage{cite}
\usepackage{url}
\usepackage{bm}
\usepackage{color}
\usepackage{subfigure}

\newcommand{\etal}{{\it et al.}~}
\newcommand{\GF}{{\mathrm{GF}}}
\newcommand{\CONSTscale}{0.27}

\definecolor{mygreen}{rgb}{0.0,0.6,0.0}
\newcommand{\revA}[1]{\textcolor{black}{#1}}    
\newcommand{\revB}[1]{\textcolor{black}{#1}}  
\newcommand{\revC}[1]{\textcolor{black}{#1}}   
\newcommand{\self}[1]{\textcolor{black}{#1}}   

\begin{document}
%
\title{Design and Performance of Rate-compatible Non-Binary LDPC Convolutional Codes}
 \author{
 \IEEEauthorblockN{Hironori Uchikawa, Kenta Kasai and Kohichi Sakaniwa}
 \IEEEauthorblockA{Dept.\ of Communications and Integrated
  Systems\\Tokyo Institute of Technology\\152-8550 Tokyo, JAPAN\\
  Email: \{uchikawa, kenta, sakaniwa\}@comm.ss.titech.ac.jp}
  }


\maketitle

\begin{abstract}
In this paper, we present a construction method of non-binary
low-density parity-check (LDPC) convolutional codes. Our construction
method is an \self{extension} of \revA{Felstr\"{o}m} and Zigangirov
construction \cite{zigangirov99} for non-binary LDPC convolutional
codes. The rate-compatibility of the non-binary convolutional code is
also discussed. \self{The} proposed rate-compatible code is designed
from one single mother (2,4)-regular non-binary LDPC convolutional code
of rate 1/2. Higher-rate codes are produced by puncturing the mother
code and lower-rate codes are produced by multiplicatively repeating the
mother code. Simulation results show that non-binary LDPC convolutional
codes \revB{of rate 1/2} 
outperform state-of-\self{the}-art binary LDPC convolutional codes
with comparable constraint bit length. Also the derived low-rate and
high-rate non-binary LDPC convolutional codes exhibit good decoding
performance without loss of large gap to the Shannon limits.
\end{abstract}


%
\IEEEpeerreviewmaketitle

\section{Introduction}
Low-density parity-check (LDPC) block codes were 
first invented by Gallager \cite{gallager-ldpc}.
Optimized binary LDPC block codes can approach very close to \revA{the} Shannon
limit with long code lengths \cite{richardson01design}.
Non-binary LDPC block codes were also invented by Gallager \cite{gallager-ldpc}. 
Davey and MacKay \cite{MacKay96lowdensity} found non-binary LDPC codes can outperform binary ones. 
Non-binary LDPC block codes have captured much attention recently due to their
decoding performance for moderate code lengths and their rate-compatibility\cite{kenta2011it}.


The convolutional counterparts of LDPC block codes, namely LDPC
convolutional codes were proposed in \cite{zigangirov99}.
LDPC convolutional codes are suitable for packet based communication
systems with variable length frames, since 
LDPC convolutional codes can be employed to
construct a family of codes of varying frame length via termination 
at both encoder and decoder.
%
Felstr\"{o}m and Zigangirov constructed the time-varying periodic LDPC
convolutional codes from LDPC block codes\cite{zigangirov99}. 
Surprisingly, the LDPC convolutional codes outperform the constituent underlying LDPC block codes. 
Recently, Kudekar \etal \revA{investigated} such decoding performance improvement by using GEXIT and 
\revA{showed} that the terminated LDPC convolutional coding increases the belief propagation (BP)
threshold, a maximum channel parameter at which decoding error
probability goes to an arbitrarily small as the code length tends to infinity, 
up to the maximum a-priori (MAP) threshold of the underlying block code\cite{kudekar2010bms} . 
\revA{In order to achieve capacity approaching performance,} LDPC convolutional codes
need to have a long constraint length, however the long constraint length 
leads to \self{long decoding latency} \cite{Papaleo2010itw}. 
The long latency is not preferred for real time communication systems.
Moreover it is desired to design rate-compatible convolutional
codes that cover from low rate to high rate, 
to establish reliable communication systems over channels with wide range of noise strength.
%

In this paper, we \self{study} a non-binary LDPC convolutional code and its
rate-compatibility.
We modify the construction method \cite{zigangirov99}, in order to
construct a non-binary (2,4)-regular LDPC convolutional code.
Using the (2,4)-regular LDPC convolutional code as a mother code, 
a rate-compatible non-binary LDPC convolutional code can be derived.
High-rate non-binary LDPC convolutional codes are produced by puncturing the mother LDPC convolutional code.
Lower-rate codes are produced by multiplicatively repeating the mother
code\cite{kenta2011it}. Simulation results show the
non-binary LDPC convolutional code of rate 1/2 outperforms binary
LDPC convolutional codes with smaller decoding latency, 
and also have good performance for rates from 1/4 to 7/8 
without large loss from the Shannon limits.


The paper is organized as follows. In Section \ref{sec:ldpc-cc}, we
introduce terminated LDPC convolutional codes over GF$(2^{p}$). Then
we give a construction method and simulation results for a mother 1/2 code
in Section \revB{\ref{sec:const}}.
Section \ref{sec:rate} explains how to produce low-rate codes 
and high-rate codes from the mother code. 
Finally, we give conclusions in Section \ref{sec:conclusions}.
\section{Terminated LDPC Convolutional Codes over GF$(2^p)$}
\label{sec:ldpc-cc}
In this section, we present a brief overview of terminated ($m_\mathrm{s},J,K$) regular
LDPC convolutional codes over GF($2^p$).

\subsection{Code Definition}
For convenience, we follow the notations in
\cite{Costello06comp} to describe time-varying syndrome former 
(transposed parity-check) matrix of LDPC convolutional codes.
An ($m_\mathrm{s},J,K$) regular LDPC convolutional code over GF($2^p$)
is the set of sequences $\bm{v}\in {\rm GF}(2^p)^{c(N+Z)}$ 
satisfying the equation $\bm{vH}^{T} = \bm{0}$, where
$Z$ is a time unit for termination.
The length of the codeword $\bm{v}$ is given as $c(N+Z)$.
A syndrome former matrix $\bm{H}^{\mathsf{T}}$ is defined as (\ref{eqn:H}). 
The submatrix $\bm{H}^{\mathsf{T}}_{i}(t)$, $i = 0, 1, \ldots, m_\mathrm{s}$, is
\self{a} $c \times (c-b)$ non-binary matrix over GF$(2^p)$ which forms 
\begin{equation}
 \bm{H}^{\mathsf{T}}_{i}(t) = \begin{bmatrix}
		      h_{i}^{(1,1)}(t) & \cdots & h_{i}^{(1,c-b)}(t) \\
		      \vdots         &        & \vdots           \\
		      h_{i}^{(c,1)}(t) & \cdots & h_{i}^{(c,c-b)}(t)
		     \end{bmatrix}, 
\end{equation}
where $h_{i}^{(\gamma,\eta)}(t) \in \mathrm{GF}(2^p), \text{ for } ~\gamma = 1,\ldots,c, ~\eta = 1,\ldots,c-b, ~p \ge 2.$
\begin{figure*}[ht] 
\revA{
 \begin{equation}
  \bm{H}^{\mathsf{T}} = 
   \begin{bmatrix}
    \bm{H}^{\mathsf{T}}_{0}(0) & \bm{H}^{\mathsf{T}}_{1}(1) & \cdots & \bm{H}^{\mathsf{T}}_{m_\mathrm{s}}(m_\mathrm{s}) \\
    & \bm{H}^{\mathsf{T}}_{0}(1) & \cdots &
    \bm{H}^{\mathsf{T}}_{m_\mathrm{s}-1}(m_\mathrm{s}) & \bm{H}^{\mathsf{T}}_{m_\mathrm{s}}(m_\mathrm{s} + 1) \\
    & & \ddots & & \ddots  &  \\
    & & \bm{H}^{\mathsf{T}}_{0}(t)  & \cdots & \cdots & \bm{H}^{\mathsf{T}}_{m_\mathrm{s}}(m_\mathrm{s} + t) & \\
    & & & \ddots & & \ddots & \ddots &  \\
    & & & & \ddots & \ddots & \ddots & \ddots &  \\
    & & & & & & \bm{H}^{\mathsf{T}}_{0}(N+Z-1) & \cdots & \bm{H}^{\mathsf{T}}_{m_\mathrm{s}}(m_\mathrm{s}+N+Z-1) 
    \label{eqn:H}
   \end{bmatrix}
 \end{equation}
}
\hrulefill 

\end{figure*}
$\bm{H}^{\mathsf{T}}_{0}(t)$ needs to be full rank for systematic encoding and
$\bm{H}^{\mathsf{T}}_{m_\mathrm{s}}(t)$ should be a nonzero matrix to
maintain a constraint length $\nu_{s} = (m_\mathrm{s} + 1) c$.
$m_\mathrm{s}$ is the maximum width of the nonzero entries in the matrix 
$\bm{H}^{\mathsf{T}}$, and is referred to as syndrome former memory, 
associated constraint bit length is defined as $\nu_{b} = (m_\mathrm{s} + 1) c p$.

In a practical manner, a syndrome former matrix has a periodical structure. 
Therefore $\bm{H}^{\mathsf{T}}_{i}(t) = \bm{H}^{\mathsf{T}}_{i}(t + T)$
is satisfied for all $t$, where $T$ is called the period of the matrix. 
%
For large $N$, the rate $R$ of this code is given as
\begin{align*}
 R=\frac{b}{c(1+Z/N)} = \frac{b}{c}\quad (N\to\infty)
\end{align*}
$\bm{H}^{\mathsf{T}}$ has $J$ nonzero entries in each row and $K$ nonzero
entries in each column, except at the first $m_\mathrm{s} (c - b)$ columns
and the last $Z$ columns.


\subsection{Encoding}
\begin{figure}[htb]
\centering
\includegraphics[width=0.47\textwidth]{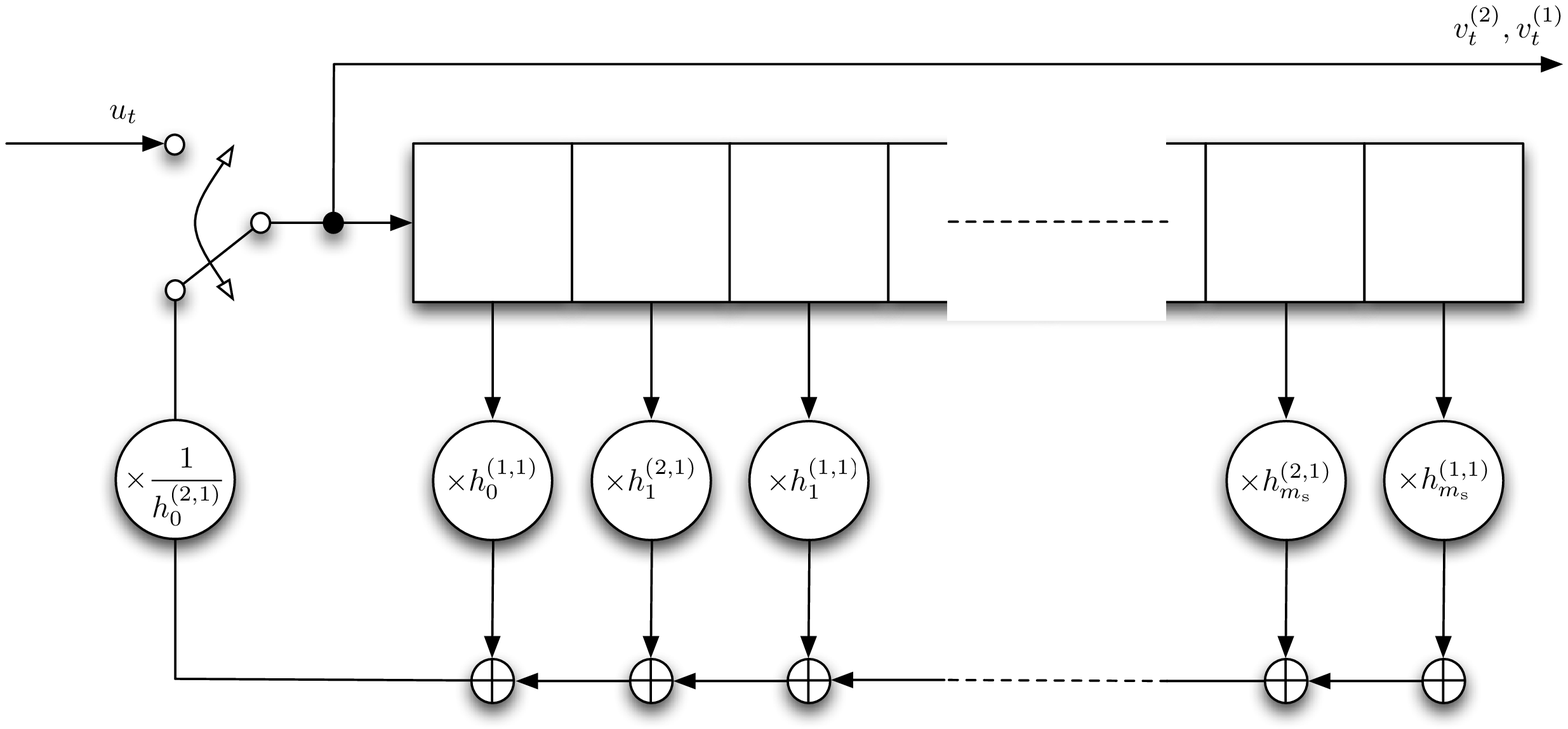}
\caption{\self{A shift register based encoder} for non-binary LDPC convolutional codes
 with R = 1/2}
\label{fig:shift_reg}
\end{figure}

Encoding of the non-binary LDPC convolutional codes is accomplished in a systematic manner.
Let $\bm{u}$ be the information sequence, where
\begin{align*}
\bm{u} &:= (\bm{u}_{0}, \bm{u}_{1}, \cdots, \bm{u}_{t},
\cdots,\bm{u}_{N+Z-1})\in \GF(2^p)^{b(N+Z)}, \\
\bm{u}_{t} &:= (u_{t}^{(1)}, \cdots, u_{t}^{(b)})\in
 \mathrm{GF}(2^{p})^b.
\end{align*}
This information sequence is encoded into the coded sequence $\bm{v}$ by a convolutional encoder, where
\begin{align*}
\bm{v} &:= (\bm{v}_{0}, \bm{v}_{1}, \cdots, \bm{v}_{t}, \cdots,\bm{v}_{N+Z-1})\in \mathrm{GF}(2^{p})^{c(N+Z)},\\
 \bm{v}_{t} &:= (v_{t}^{(1)}, \cdots, v_{t}^{(c)}) \in \mathrm{GF}(2^{p})^c.
\end{align*}
The coded sequence satisfies 
$\bm{vH}^{T} = \bm{0}$ which can be rewritten as 
\begin{align}
 \sum_{i=0}^{t} \bm{v}_{t-i}\bm{H}_{i}^{T}(\revA{t}) &= 0, ~\text{for} ~0 \leq t < m_\mathrm{s},
 \label{eqn:check1} \\
 \sum_{i=0}^{m_\mathrm{s}} \bm{v}_{t-i}\bm{H}_{i}^{T}(\revA{t}) &= 0, ~\text{for} ~m_\mathrm{s} \leq t \leq N+Z-1.
 \label{eqn:check2}
\end{align}
To obtain a systematic non-binary LDPC convolutional code, the last $(c-b)$ rows of
$\bm{H}_{0}^{T}(t)$ are chosen \self{so as to be a} $(c-b)\times(c-b)$ diagonal
matrix \cite{pusane06icc}. 
The code sequence $\bm{v}$ can be calculated using
Eqs. (\ref{eqn:check1}), (\ref{eqn:check2}) by
the expressions
\revB{
\begin{align*}
 v_{t}^{(j)} =& ~u_{t}^{(j)} , ~\text{for} ~j = 1, \cdots, b, \\
 v_{t}^{(j)} =& ~\frac{\sum_{k=1}^{b} v_{t}^{(k)}
 h_{0}^{(k,j-b)}(t) +
 \sum_{i=1}^{m_\mathrm{s}}\sum_{k=1}^{c}v_{t-i}^{(k)}h_{i}^{(k,j-b)}(t)
 }{h_{0}^{(j,j-b)}(t)} , \\
  & ~\text{for} ~j=b+1, \cdots, c.
\end{align*}
}
This can be easily implemented with \self{shift registers}. 
For example, the encoder of a non-binary LDPC convolutional code with $R = b/c = 1/2$ is depicted in Fig.~\ref{fig:shift_reg}. 
The number of required memory bits is equal to 
$((m_\mathrm{s} c)+b)p$ and the average complexity to
encode one parity symbol is proportional to $K-1$.
The encoding complexity is independent of the codeword length
and the syndrome former memory $m_\mathrm{s}$.
\revA{A straightforward encoder for a length $N$ non-binary LDPC block code
has a complexity per parity bit of $O(N)$, }
since the encoder multiplies the information sequence by the generator
matrix. Therefore the non-binary LDPC convolutional codes have  
a significant advantage compared to non-binary LDPC block codes in 
terms of encoding complexity.

\subsection{Decoding}
\begin{figure*}[htb]
\centering
\includegraphics[width=0.8\textwidth]{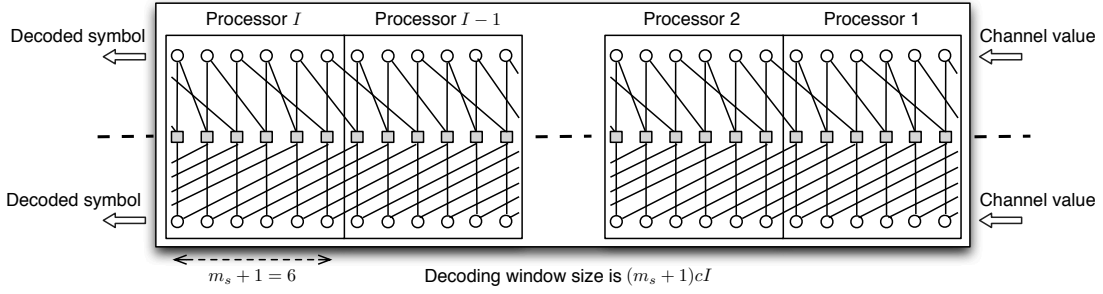}
\caption{A sliding \self{windowed} decoder for (5, 2, 4) non-binary LDPC convolutional code
 with R = 1/2}
\label{fig:swd_nb}
\end{figure*}

Decoding of the non-binary LDPC convolutional codes can be 
performed in several ways. 
A message passing algorithm similar to that for 
non-binary LDPC block codes is applicable, since the non-binary 
LDPC convolutional codes discussed in this paper are terminated.
However we have a special algorithm called sliding 
\self{windowed} decoding for the non-binary LDPC convolutional codes
\cite{Papaleo2010itw}.
Due to the convolutional structure, the distance between 
two variable nodes that are connected 
to the same check node is limited by the memory of the code.
This property can be used in order to perform 
continuous decoding of the received sequence through a
window that slides along the sequence, analogous to the
Viterbi decoder with finite path memory.
Since the sliding \self{windowed} decoder does not need 
message memory for the entire code sequence, 
it has the advantage compared to the decoder of 
the LDPC block codes in terms of decoder complexity, 

Moreover the decoding of two variable nodes 
that are at least $(m_\mathrm{s} + 1)$
time units apart can be performed independently, 
since the corresponding symbols cannot be involved in
the same parity-check equations. This indicates
the possibility of parallelizing the iterations
of the message passing decoder, through several processors
working in different regions of the Tanner graph. A pipeline
decoder based on this idea was proposed in \cite{zigangirov99}.
Figure \ref{fig:swd_nb} shows a sliding \self{windowed}
decoder for (5, 2, 4) non-binary LDPC convolutional code
for \revA{an} example.
%
The decoding time for each symbol in the decoding window is proportional
to $(m_\mathrm{s} + 1)cI$, where $I$ represents the number of the stage
of the pipeline decoder involved in the sliding window.
Intuitively, large $m_\mathrm{s}$ leads to  better performance, however the decoding latency increases with $m_\mathrm{s}$. 
It is known that non-binary LDPC block codes exhibit good \revA{decoding performance} at moderate code length. 
Therefore, it is expected that non-binary LDPC convolutional codes have good performance with small $m_\mathrm{s}$.

 \section{Construction and Performance of Rate 1/2 Non-binary LDPC Convolutional Codes}
 \label{sec:const}
 \subsection{Syndrome Former Matrix Construction}
In this section, we propose a method for constructing syndrome former
matrix $H^\mathsf{T}$ of the mother non-binary LDPC convolutional code. 
For simplicity, we concentrate on non-binary LDPC convolutional code of rate 1/2 and syndrome former matrix period $T = m_\mathrm{s}+1$.
The proposed method is easily extended to any non-binary LDPC convolutional codes of rate $R = b/c$ with
$b, c \in \revA{\mathbb{N}}$.

Felstr\"{o}m and Zigangirov \cite{zigangirov99} first introduced a syndrome former matrix
construction from a regular matrix of an LDPC block code \cite{zigangirov99}.
%
Motivated by the construction \cite{zigangirov99}, we construct 
a syndrome former matrix $H^\mathsf{T}$ of period one from a base matrix $\bm{B}^{\mathsf{T}}$ which forms 
\begin{equation}
 \bm{B}^{\mathsf{T}} = \begin{bmatrix}
			\bm{B}^{\mathsf{T}}_{0,0} & \cdots & \bm{B}^{\mathsf{T}}_{0,m_\mathrm{s}} \\
			\vdots               & \bm{B}^{\mathsf{T}}_{l,r}       & \vdots           \\
			\bm{B}^{\mathsf{T}}_{m_\mathrm{s},0} & \cdots & \bm{B}^{\mathsf{T}}_{m_\mathrm{s},m_\mathrm{s}} 
		       \end{bmatrix},
\end{equation}
where $\bm{B}^{\mathsf{T}}_{l,r}$ is size $c \times (c-b) = 2 \times 1$.
The size of the base matrix $\bm{B}^{\mathsf{T}}$ is $c(m_\mathrm{s}+1)
\times (c-b)(m_\mathrm{s}+1) = 2(m_\mathrm{s}+1) \times
(m_\mathrm{s}+1)$.  The base matrix $\bm{B}^{\mathsf{T}}$ is constructed
as follows.  First, set $\bm{B}^{\mathsf{T}}_{l,r}=[1 1]^{\mathsf{T}}$
for $l=r$ and $\bm{B}^{\mathsf{T}}_{l,r}=[0 1]^{\mathsf{T}}$ for $r=
l-1\mod (m_\mathrm{s}-1)$.  Next, put $[0 1]^{\mathsf{T}}$ or $[1
0]^{\mathsf{T}}$ at the rest of the entry positions of
$\bm{B}^{\mathsf{T}}_{l,r}$ so that each row and column of the base
matrix $\bm{B}^{\mathsf{T}}$ has weight $J$ and $K$, respectively.
\revB{In this step, the positions are chosen uniformly random by
avoiding cycles of length 4.  Replace ones in $B^\mathsf{T}$ with
randomly chosen nonzero values $\beta_{i}^{(\gamma,\eta)}(t) \in
\mathrm{GF}(2^p)\backslash \{0\}$, so that each column, i.e., check node
does not have same nonzero values.  One can further improve the error
floors by choosing nonzero values by the methods developed in
\cite{Poulliat08design}\cite{savin2008isabel}.}

In order to obtain the syndrome former matrix $H^\mathsf{T}$ of period
one, 
cut the base matrix $\bm{B}^{\mathsf{T}}$ along the diagonal,
and the lower diagonal part is appended to the right side of 
the upper diagonal part.
The resulting diagonal shaped matrix $\hat{\bm{B}}^{\mathsf{T}}$
forms a syndrome former matrix $H^\mathsf{T}$ of period one as follows. 

\begin{align}
 \hat{\bm{B}}^{\mathsf{T}} &= 
  \begin{bmatrix}
   \bm{B}^{\mathsf{T}}_{0,0}\hspace{3mm} & \cdots  & \bm{B}^{\mathsf{T}}_{0,m_\mathrm{s}} &\\
   & \ddots     & & \ddots         \\
   & & \bm{B}^{\mathsf{T}}_{m_\mathrm{s},m_\mathrm{s}}\hspace{3mm}&
   \cdots & \bm{B}^{\mathsf{T}}_{m_\mathrm{s},m_\mathrm{s}-1} 
  \end{bmatrix} \\
 &= 
  \begin{bmatrix}
   \bm{H}^{\mathsf{T}}_{0}(0) & \cdots & \bm{H}^{\mathsf{T}}_{m_\mathrm{s}}(m_\mathrm{s}) \\
   & \ddots & & \ddots  \\
   & & \bm{H}^{\mathsf{T}}_{0}(m_\mathrm{s})  & \cdots &
   \bm{H}^{\mathsf{T}}_{m_\mathrm{s}}(2 m_\mathrm{s}) 
  \end{bmatrix}
\end{align}

By stacking $ \hat{\bm{B}}^{\mathsf{T}}$ to achieve a desired size, 
we obtain a syndrome former matrix $H^\mathsf{T}$ as in \eqref{eqn:H}.

\begin{figure*}[ht]
  \begin{center}
   \subfigure[ ]{ 
   \includegraphics[scale=\CONSTscale]{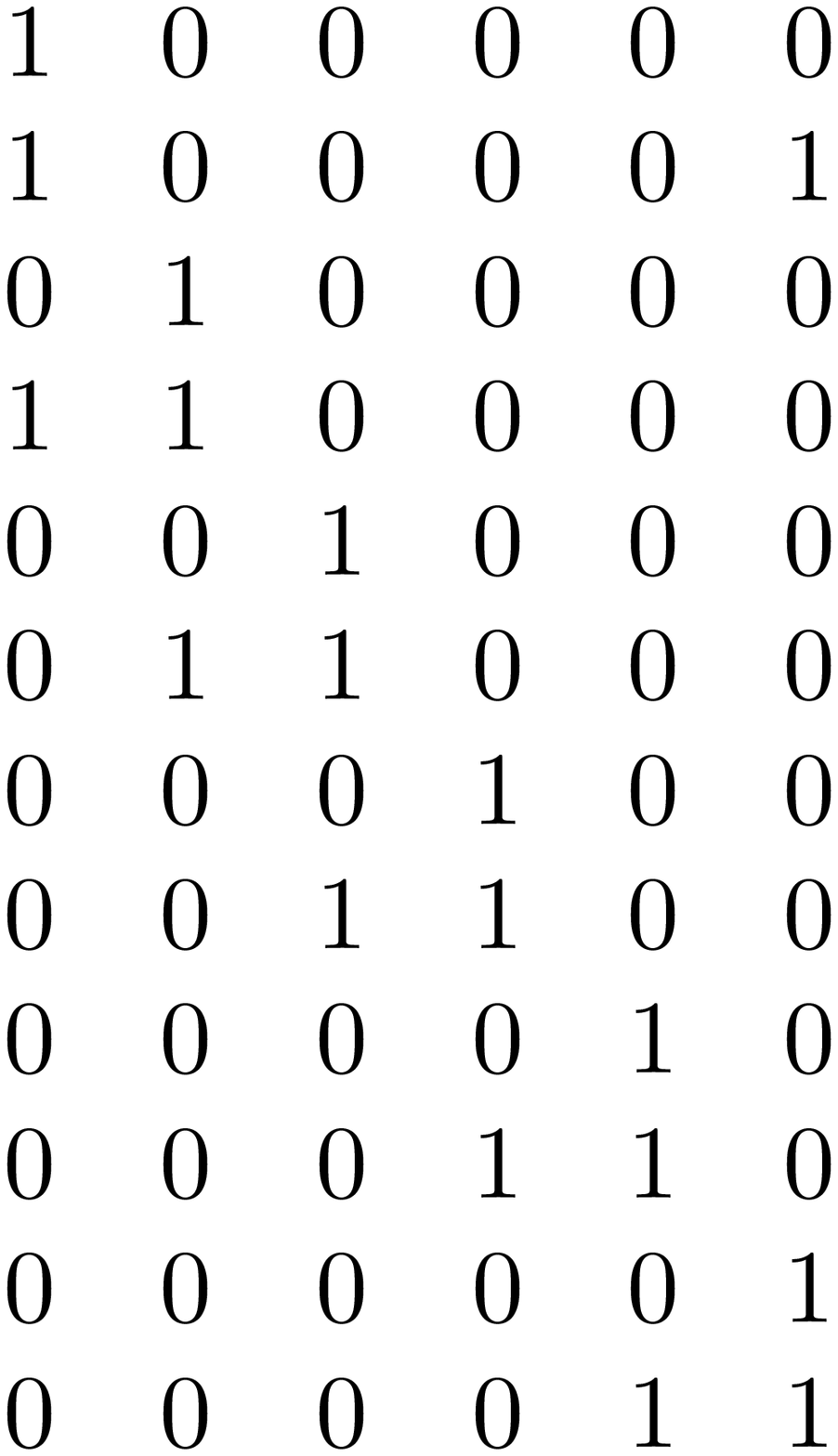}
   \label{fig:our_const1}}
   \subfigure[ ]{
   \includegraphics[scale=\CONSTscale]{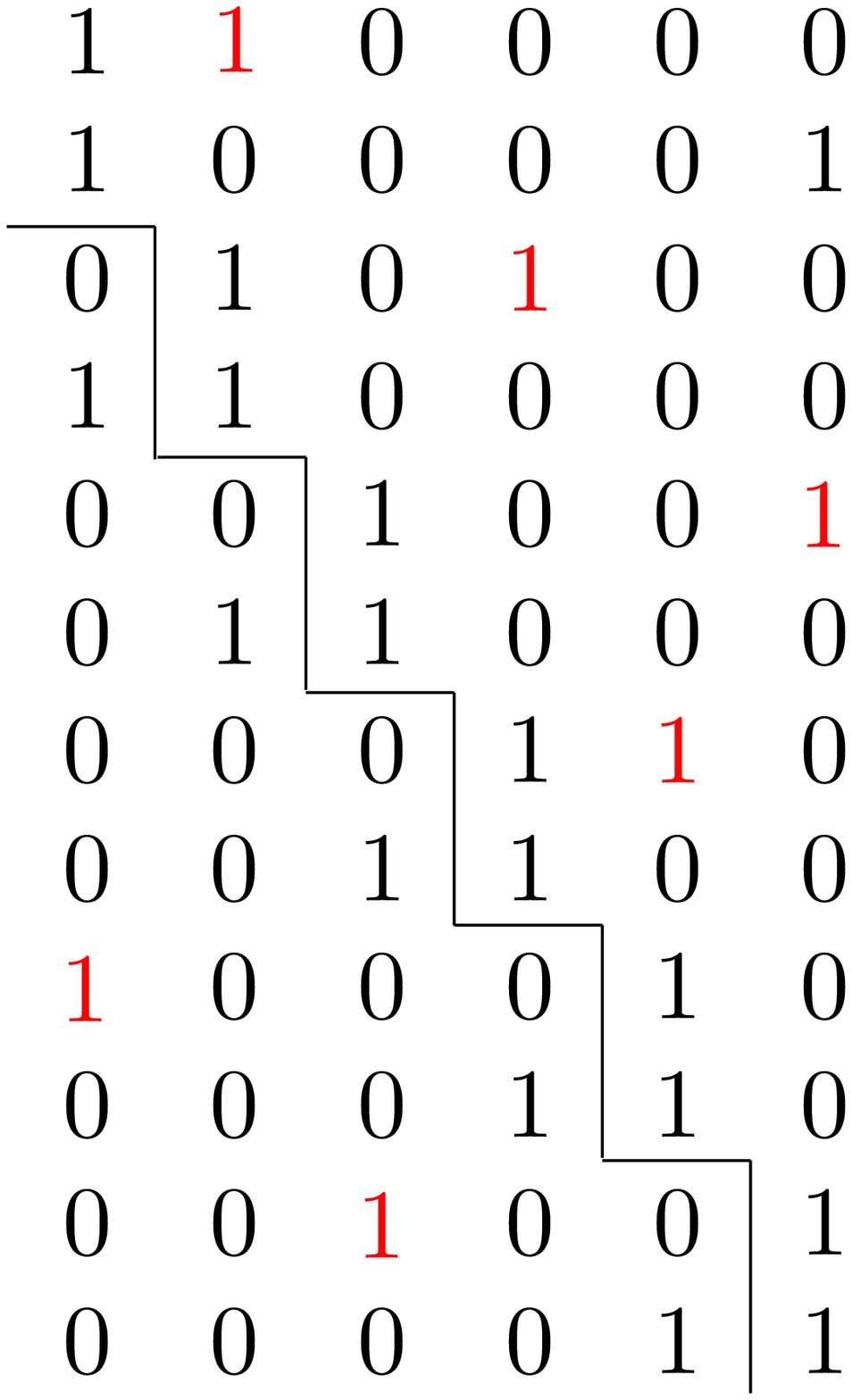}
   \label{fig:our_const2}}
   \subfigure[ ]{
   \includegraphics[scale=\CONSTscale]{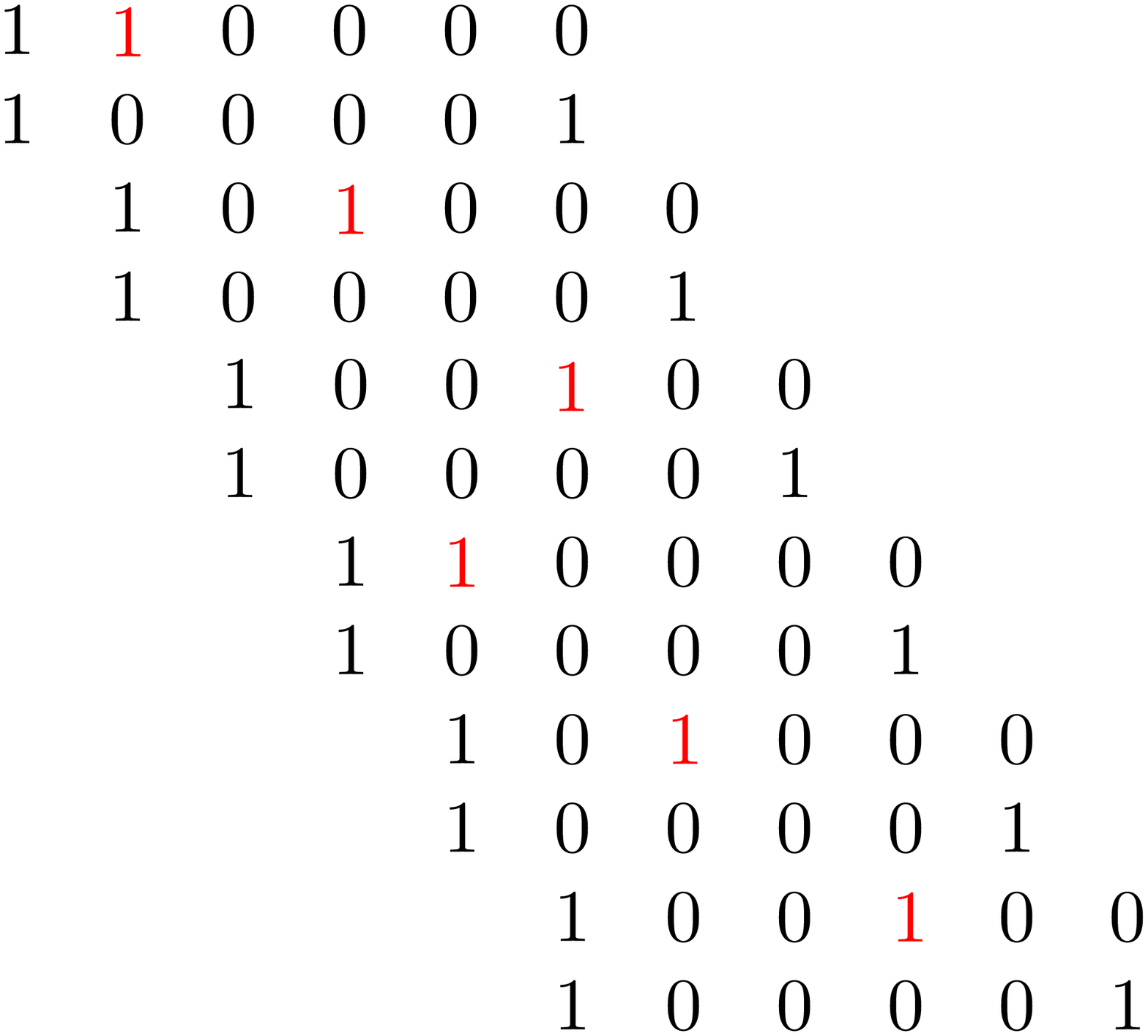}
   \label{fig:our_const3}}
   \subfigure[ ]{
   \includegraphics[scale=\CONSTscale]{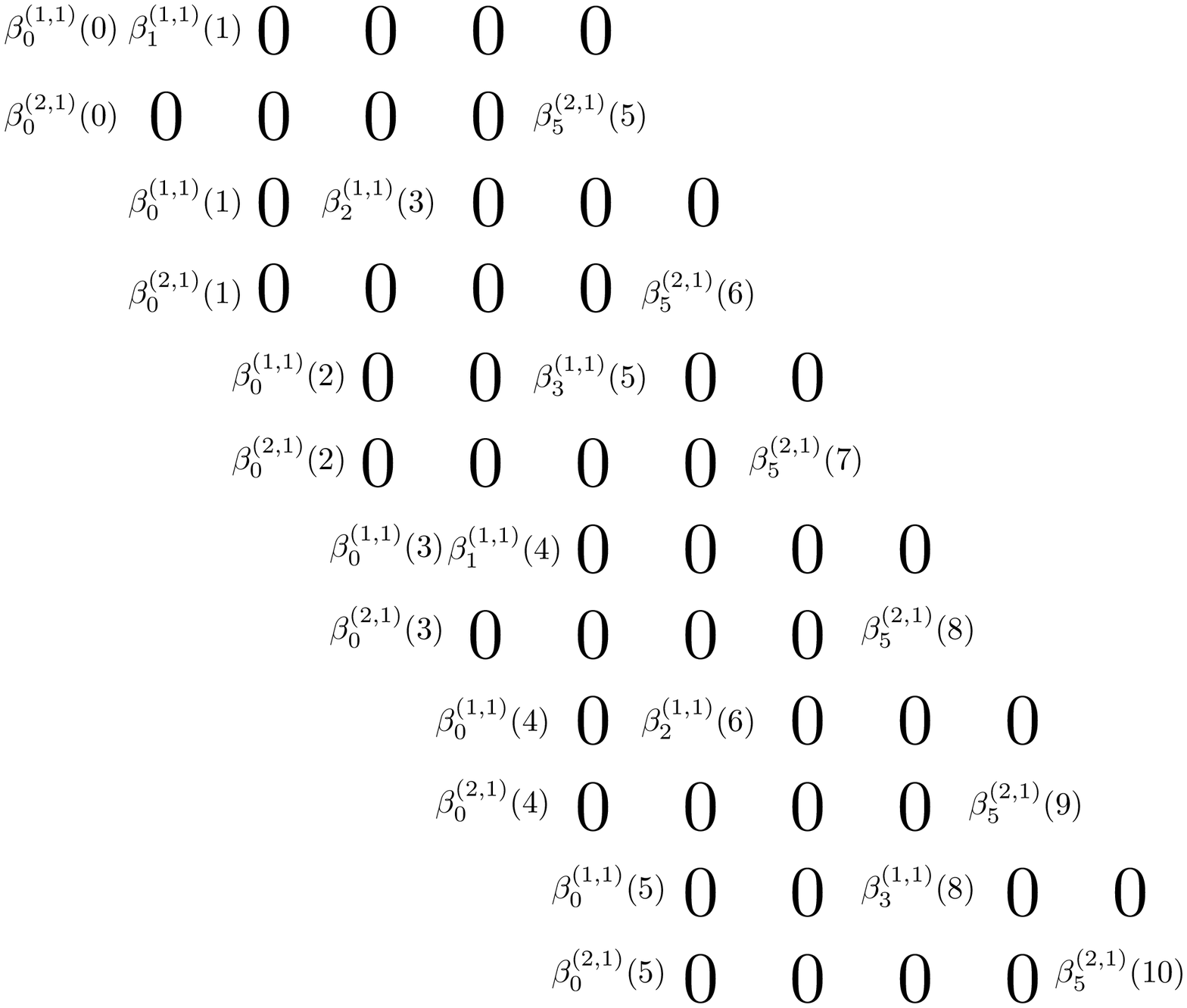}
   \label{fig:our_const4}}
   \subfigure[ ]{
   \includegraphics[scale=\CONSTscale]{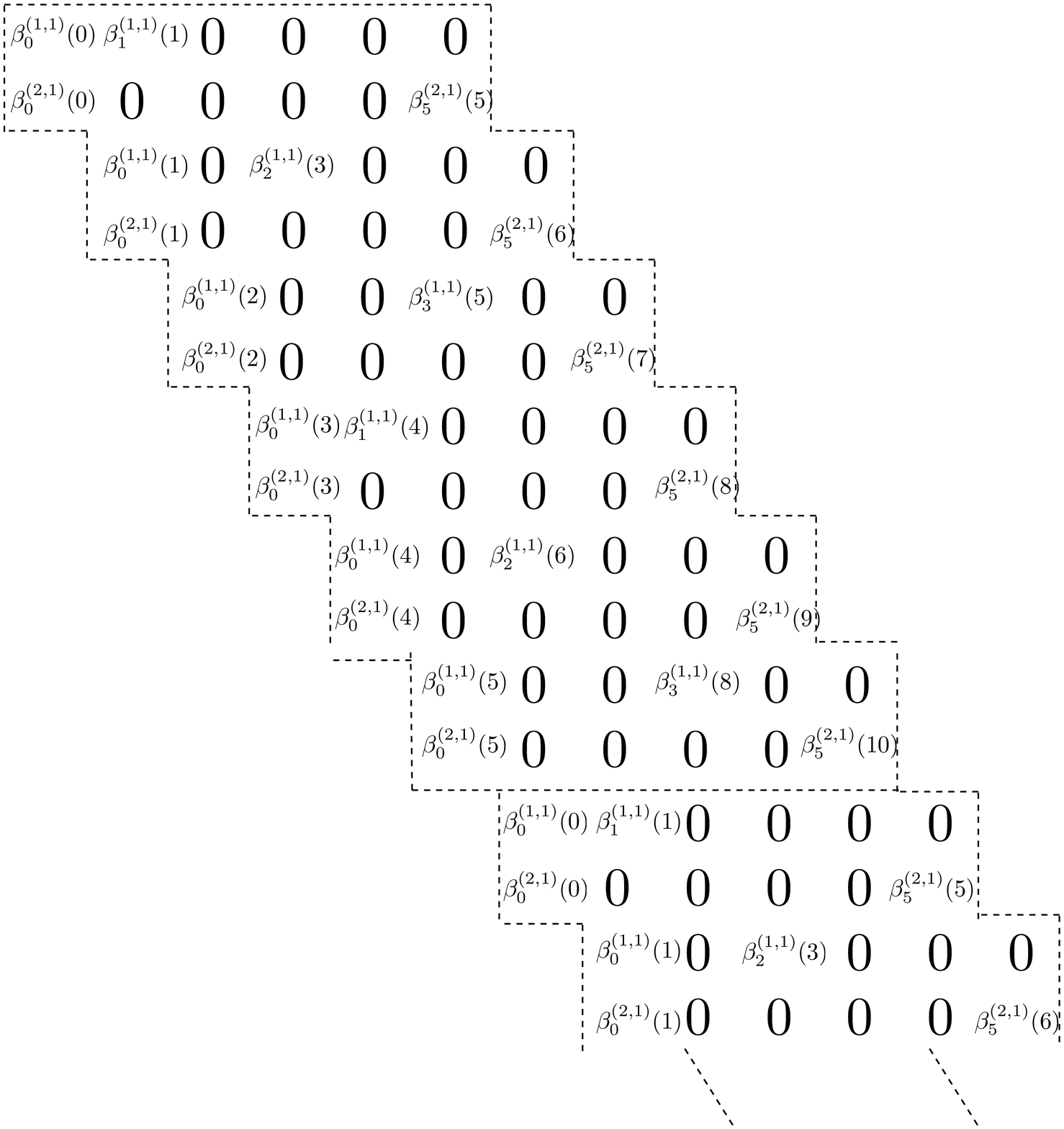}
   \label{fig:our_const6}}
   \caption{Construction procedure of a (5,2,4)-regular non-binary LDPC convolutional code.}
   \label{fig:our_const}
  \end{center}
 \end{figure*}

\begin{figure}[htb]
\centering
\includegraphics[width=8.7cm]{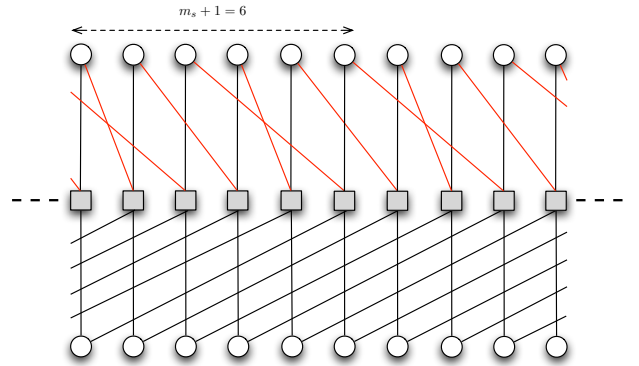}
\caption{Tanner graph of (5,2,4)-regular LDPC convolutional code}
\label{fig:2_4_cc_graph}
\end{figure}

We give an example of the syndrome former matrix construction for 
a (5, 2, 4)-regular non-binary LDPC convolutional code. 
\self{Figure} \ref{fig:our_const} shows the construction procedure. 
We first put the matrices $[1 1]^{\mathsf{T}}$ and $[0 1]^{\mathsf{T}}$
on the base matrix of size $12 \times 6$ shown in Fig.~\ref{fig:our_const1}. 
In the next step, we put ones randomly on odd rows of the matrices 
\revC{without cycles of length 4} so that 
each column has weight 4. Ones placed in this step is colored 
red in Fig.~\ref{fig:our_const2} and the diagonal shaped matrix is 
shown in Fig.~\ref{fig:our_const3}. The corresponding Tanner 
graph is shown in Fig.~\ref{fig:2_4_cc_graph}.
The upper circle nodes in Fig.~\ref{fig:2_4_cc_graph} represent 
the odd rows and the lower circle nodes represent the even rows 
on the base matrices. Ones placed randomly correspond to the 
connection of red edges in Fig.~\ref{fig:2_4_cc_graph}.
The size of the code ensemble, i.e.,  $(m_\mathrm{s}, 2,4)$-regular matrices
is given as $(m_\mathrm{s}+1)!$. 
Then we replace ones with nonzero elements 
\revC{ $\beta_{i}^{(\gamma,\eta)}(t) \in \mathrm{GF}(2^p)\backslash
\{0\}$ }
(see Fig.~\ref{fig:our_const4})\revC{, so that
each column does not have same nonzero values.}
In the final step, 
the diagonal shaped matrix 
is repeated periodically in order to achieve the desired size syndrome former 
matrix of an LDPC convolutional code (see Fig.~\ref{fig:our_const6}).

One might think this Tanner graph is too structured and lacks of randomness. 
However the edge coefficients are randomly chosen so that 
the equivalent binary representation of the code has a large degree of
freedom. In fact, the size of the code ensemble with such non-binary syndrome
former matrices is given as $(m_\mathrm{s}+1)!\times(2^p-1)^{4(m_\mathrm{s}+1)}$.

\begin{figure}[t]
\centering
\includegraphics[width=0.5\textwidth]{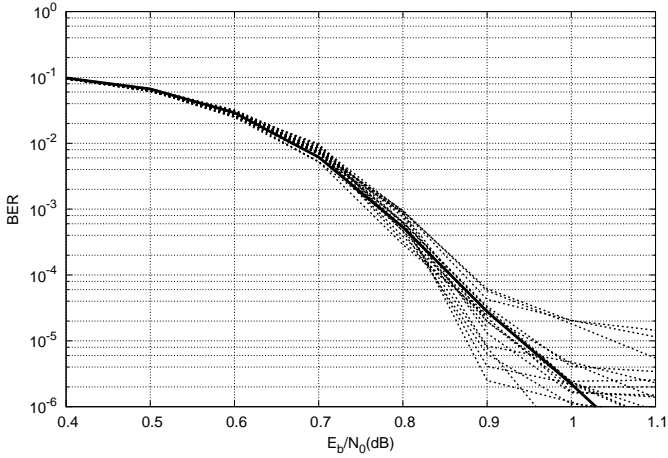}
\caption{
\revC{Bit error rate (BER) of 20 instances of (52, 2, 4) LDPC 
convolutional code over GF$(2^8)$. 
The code of the solid line is used in the following section.
We observe that the average of the error floor performance is
between $10^{-5}$ and $10^{-6}$.
}
}
\label{fig:52.2.4}
\end{figure}

\revC{Figure \ref{fig:52.2.4} shows the bit error rate (BER) curves 
for 20 random samples of (52, 2, 4) LDPC convolutional code over GF$(2^8)$.
We observe that the curves have large deviation below the BER of $10^{-4}$
because of the dispersion of the error floor performance.
Since the BER curve of the code is the average of 20 instances,
we employ the code of the solid line in the following section.
Also we believe that generating 20 instances is enough to obtain
the average error performance code by using our construction method.
}

\begin{figure}[t]
\centering
\includegraphics[width=0.5\textwidth]{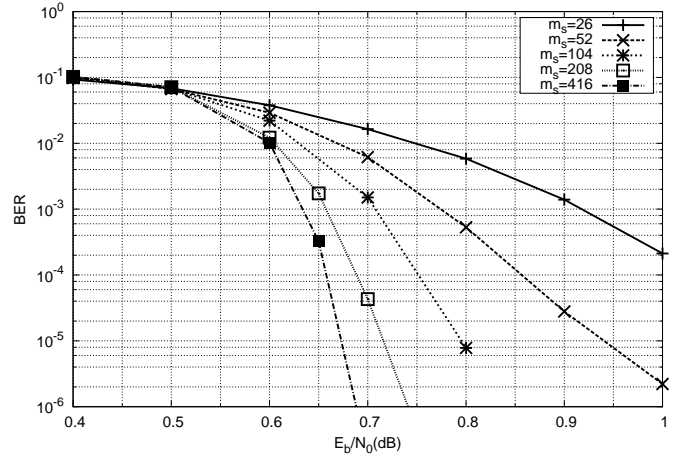}
\caption{
\revC{The BER performances of 
$m_\mathrm{s}=26, 52, 104, 208,$ and $416$, $J=2, K=4$ non-binary
LDPC convolutional code over GF$(2^8)$. All of these codes
are rate 1/2. Error correction performance improves 
with increasing $m_\mathrm{s}$.}
}
\label{fig:var_ms}
\end{figure}

\revB{In general, the binary LDPC convolutional codes with large 
$m_\mathrm{s}$, i.e., large $\nu_{b}$ 
have good error correction performance. 
The binary LDPC convolutional codes with $\nu_{b} > 2000$ 
were discussed in \cite{Costello06comp}\cite{pusane06icc}.
From Fig. \ref{fig:var_ms}, we can claim the same statement 
for the non-binary LDPC convolutional codes.
In the point of view of the error correcting performance, 
large $\nu_{b}$ is preferred, however
we expect that such codes have large decoding latency.
In order to show the superior performance of our proposed codes, 
we will employ the non-binary LDPC convolutional codes
with small $\nu_{b} = 848$, i.e., $m_\mathrm{s}=52$ in this paper.
Simulation results show that the non-binary LDPC convolutional codes
have good error correction performance, nevertheless such small
$\nu_{b}$.}

\subsection{Simulation Results}
\label{sec:sim}
In this section, we compare the non-binary LDPC convolutional codes
with binary LDPC convolutional codes and non-binary LDPC block codes.

The transmissions over the AWGN channel with BPSK are assumed.
The sum-product algorithm using the fast Fourier transform (FFT)
is employed for decoding.
The number of iterations is set to 50. 

\begin{figure}[!ht]
\centering
\includegraphics[width=0.5\textwidth]{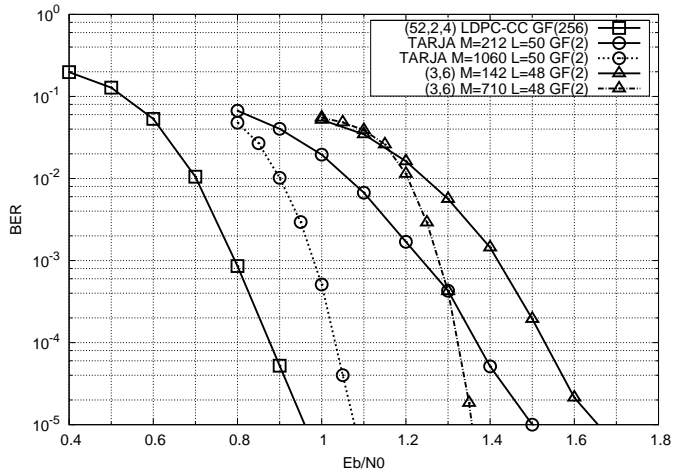}
\caption{Simulation results of a (52, 2, 4) LDPC convolutional 
code over GF$(2^8)$ (square), a binary TARJA convolutional code
 \cite{lentmaier10turbo} (circle), 
and a binary (3,6) convolutional code (triangle).
Both binary codes are constructed with equivalent constraint 
bit length $\nu_{b} = 848$ (solid lines) to the non-binary code
and 5 times longer constraint bit length (dashed lines). 
All of these codes are of rate 1/2.
}
\label{fig:NB24vsB36}
\end{figure}

In Fig. \ref{fig:NB24vsB36}, we compare the BER of
a (52, 2, 4) LDPC convolutional code over GF$(2^8)$ with binary 
LDPC convolutional codes.
In the simulation, the termination time unit $Z$ is set to $m_{s}$.
Therefore the termination bit length of the (52, 2, 4) LDPC 
convolutional code over GF$(2^8)$ is 832. The encoded information 
bit length $bpN$ is set to 40000. The resulting code rate 
is almost 0.495.

For the binary convolutional codes, one is a (3,6) regular LDPC convolutional
code \cite{Papaleo2010itw} and the other is a terminated 
\revA{accumulate-repeat-jagged-accumulate} (TARJA)
convolutional code \cite{lentmaier10turbo}. Both codes are expanded
from their protograph with random permutation matrices of sizes $M=142$
and 212, making
their constraint bit lengths equivalent to the non-binary code.
Both binary LDPC convolutional codes with 5 times
longer constraint bit length are also shown for comparison.
Termination factor $L$ is set to make their code rates equivalent
to the rate of the non-binary code.
It is observed that
the (52, 2, 4) non-binary LDPC convolutional code provides superior
performance (about 0.3 dB at a BER of $10^{-4}$)
with smaller decoding latency to the state-of-\self{the}-art binary TARJA convolutional code
($M=1060, L=50$). \revB{Also it can be seen that the non-binary code does not have error
floors down to BER $10^{-5}$. }

\begin{figure}[ht]
\centering
\includegraphics[width=8.7cm]{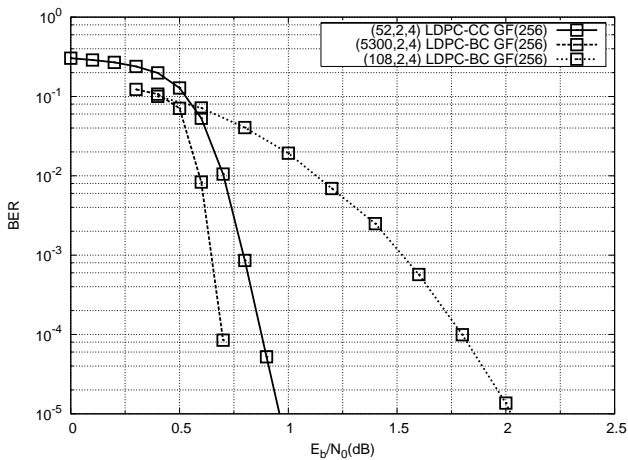}
\caption{Simulation results of a (52, 2, 4) LDPC convolutional 
code (CC) (solid lines) with same memory requirements 
and processor complexity, $N=5300$ and 108 respectively, 
LDPC block codes (BC) (dashed lines). 
All of these codes are of rate 1/2 and defined over GF$(2^8)$.
}
\label{fig:NBCCvsNBBC}
\end{figure}

Figure \ref{fig:NBCCvsNBBC} shows the performances of two (2, 4) LDPC 
block codes over GF$(2^8)$, which have \self{the symbol nodes of degree 2 and
the check nodes of the degree 4}. The block lengths were chosen so that in
one case the decoders have the same processor complexity \cite{Costello06comp}, i.e., 
$N = \nu_{s}$, and in the other case the same memory requirements, 
i.e., $N = \nu_{s} \cdot I$. 
For the same processor complexity, the convolutional code outperforms
the block code by about 0.9 dB at a BER of $10^{-4}$. However the
block code outperforms the convolutional code by about 0.15 dB at a
BER of $10^{-4}$ for the same memory requirements. 
In binary cases, the convolutional code slightly outperforms the block
code for the same memory requirements \cite{Costello06comp}.  However
\self{this is not the case of non-binary codes} in our simulation result.
We will discuss this phenomenon in the next section.


\subsection{Discussion and BP threshold analysis}
\label{sec:bp}
\begin{figure}[t]
\centering
\includegraphics[width=0.5\textwidth]{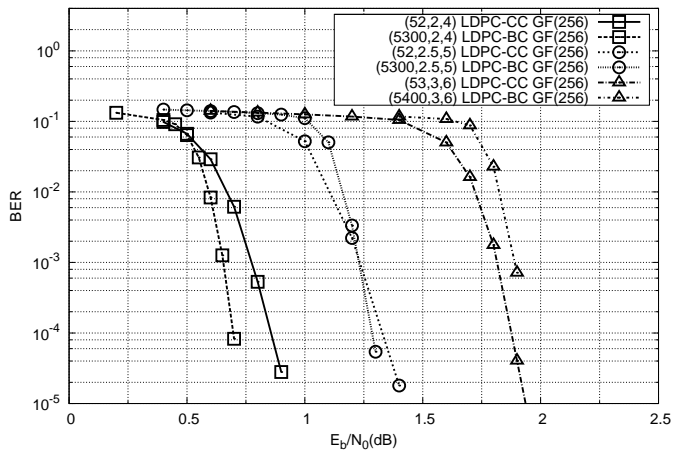}
\caption{Simulation \self{results} of $(m_\mathrm{s}, J, K)$ LDPC convolutional codes (CCs)
of different degrees of symbol nodes and $(N, J, K)$ block codes (BCs) with
comparable memory requirements. 
All of these codes are of rate 1/2 and defined over GF$(2^8)$.
The performance of the CC becomes better than that of the BC with
increasing $d_{v}$.}
\label{fig:NB24_2p55_36}
\end{figure}

\revA{In order to explain the reason why the bit error rate 
performance of the non-binary LDPC convolutional codes
is worse than the corresponding block codes in the
previous section, we show another simulation result and also
numerical calculation of the BP threshold.
In Fig. \ref{fig:NB24_2p55_36}, we
provide simulation results of non-binary LDPC convolutional codes of
different degrees of symbol nodes and block codes with comparable memory
requirements.
With an abuse of notation, 
a (\revB{52}, 2.5, 5) convolutional code and a (5300, 2.5, 5) block code have
equivalent number of symbol nodes of degree 2 and 3, so that the
average degrees of symbol nodes are 2.5. In other words, the 
syndrome former matrix of $J = 2.5$ convolutional codes has 
equivalent number of row weight 2 and 3 rows.}
In Fig. \ref{fig:NB24_2p55_36}, it can be seen that the non-binary LDPC
convolutional code with $J = 3$ outperforms the block codes like the
binary case \cite{Costello06comp}. On the other hand, the non-binary
LDPC convolutional code with $J=2$ have slightly worse performance than
the corresponding block codes. 
Lentmaier \etal describes that BP 
thresholds of regular LDPC convolutional codes improve
\self{by} increasing $J$ in \cite{Lentmaier07iterative}. Kudekar \etal
\revA{investigated} such decoding performance improvement by using GEXIT and
\revA{showed} that the LDPC convolutional coding increases the BP
threshold \revA{up to}
the MAP threshold of the underlying block code
\cite{kudekar2010bms}. 
\revA{Kudekar \etal called this phenomenon 
{\it threshold saturation} \cite{kudekar2011it}.
From the above discussion, we consider that 
the BP threshold of the $J = 2$ non-binary LDPC block code
is already very close to its MAP threshold, so that the corresponding
convolutional code cannot outperform in the simulation.
In order to verify the consideration, we will compute the BP threshold.}

Since density evolution over the AWGN channel for non-binary
LDPC codes with large field size becomes
computationally intensive and tractable only for the BEC, 
we will calculate the BP thresholds over BEC by using density
evolution for the non-binary LDPC
code ensembles with parity-check matrices defined over the general
linear group ${\rm GL}(\GF(2), p)$ \cite{rathi05iee}, instead of Galois field.
This is a fair approximation, since in \cite{rathi05iee}, it is 
reported that the threshold for the code ensemble with parity-check matrices
defined over GF($2^p$) and ${\rm GL}(\GF(2), p)$ have almost the same thresholds
within the order of $10^{-4}$.
We follow an ensemble representation of non-binary LDPC convolutional
codes in \cite{kudekar2011it} for density evolution.
\self{The} BP thresholds of non-binary LDPC convolutional and block codes
over binary erasure channel (BEC) are shown in Table \ref{tab:ebpth}.
It can be observed that 
both BP thresholds of (2,4) regular LDPC convolutional and 
block codes are almost same at the identical $p$. 
On the other hand, a BP threshold of (3,6) regular
LDPC convolutional codes is increasing with increasing $p$, however 
that of block codes is decreasing. 
This result implies that it is easier to see threshold saturation for
the (3,6) regular LDPC convolutional codes at a moderate length than the
(2,4) regular LDPC convolutional codes.
Since the BP threshold of (2,4)
regular LDPC convolutional codes is slightly higher than that of the
block code at the large $p$, we believe the threshold saturation could
be observed with sufficiently large lengths.

\begin{table}[t]
\caption{BP threshold values of (2,4) and (3,6) regular LDPC convolutional
 codes (CC) and block codes (BC) \self{over ${\rm GL}(\GF(2), p)$}. 
Coupling factors $L$ \cite{kudekar2011it} of CC is 64.}
\label{tab:ebpth}
\begin{center}
  \begin{tabular}{ccccc}
   $p$ & (2,4) CC & (2,4) BC & (3,6) CC & (3,6) BC \\\hline
   \self{1} &0.333333 &0.333333  &0.4881    & 0.4294	 \\
   \self{2} &0.409912 &0.409604  &0.490723  & 0.423472 \\
   \self{3} &0.453491 &0.450595  &0.49353   & 0.412203 \\
   \self{4} &0.474976 &0.468011  &0.494629  & 0.398902 \\
   \self{5} &0.48584  &0.474147  &0.496094  & 0.385472 \\
   \self{6} &0.490234 &0.47464   &          &  \\
 \end{tabular}
\end{center}
\end{table}

\section{Rate-Compatibility of Non-binary \\LDPC Convolutional Codes}
\label{sec:rate}
In this section, we discuss rate-compatibility of non-binary LDPC convolutional codes.
Rate-compatible non-binary LDPC convolutional codes are defined over GF$(2^8)$ in
this section for convenience. 
For non-binary LDPC block codes, puncturing and multiplicative repetition give good
rate-compatibility \cite{kenta2011it}.
We show that those techniques are also applicable in non-binary LDPC
convolutional codes.

\begin{figure}[h]
\centering
 \includegraphics[width=0.48\textwidth]{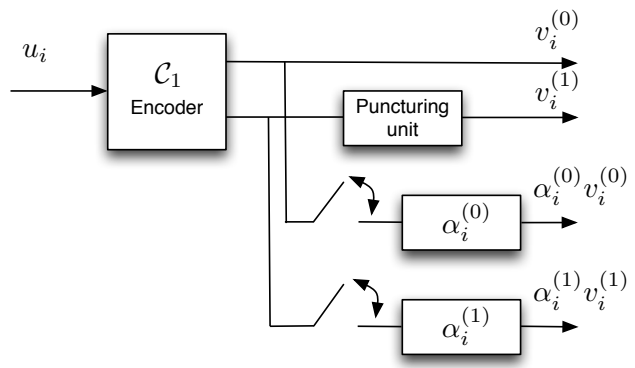}
\caption{\self{An} encoder structure of the proposed rate-compatible non-binary LDPC
 convolutional code}
\label{fig:encoder}
\end{figure}

An encoder structure of a rate-compatible non-binary LDPC convolutional code
\self{is} shown in Fig.~\ref{fig:encoder}. This encoder is composed of an encoder
of a mother code ${\cal C}_{1}$, a puncturing unit and multiplicative repeaters.
Coefficients $\revB{\alpha_{t}^{(1)}, \alpha_{t}^{(2)}}, t = 0,\ldots,N+Z-1$ are chosen
randomly from GF$(2^8)\backslash \{0,1\}$. The mother code ${\cal C}_{1}$
is the non-binary LDPC convolutional code discussed in the previous section, and the encoding 
process is accomplished with \self{shift registers}.
\self{The information symbols $u_{t} \in \mathrm{GF}(2^8)$  enter the encoder.}
The corresponding encoded symbols of the ${\cal C}_{1}$ encoder are given by
$(v_{t}^{(1)}, v_{t}^{(2)})$. 
The encoder is systematic, i.e.,  $v_{t}^{(1)} = u_{t}$.

\revA{By puncturing the parity symbols $v_{t}^{(2)}$ for $t =0,\ldots,N+Z-1$, the coding rate increases. 
Some puncturing patterns used in the simulation are shown in Table \ref{tb:punc_pat}.
On the other hand, by multiplicatively repeating the encoded symbols $(v_{t}^{(1)}, v_{t}^{(2)})$ with 
multiplicative repetition coefficient $(\revB{\alpha_{t}^{(1)}}v_{t}^{(2)}, \revB{\alpha_{t}^{(1)}}v_{t}^{(2)})$, the coding rate decreases down to 1/4.
The more we increase multiplicatively repeated symbols, the more
overall rate decreases. }
We can also design various \self{rates} with combining puncturing and multiplicative repetition.  

\begin{figure}[h]
\centering
 \includegraphics[width=0.5\textwidth]{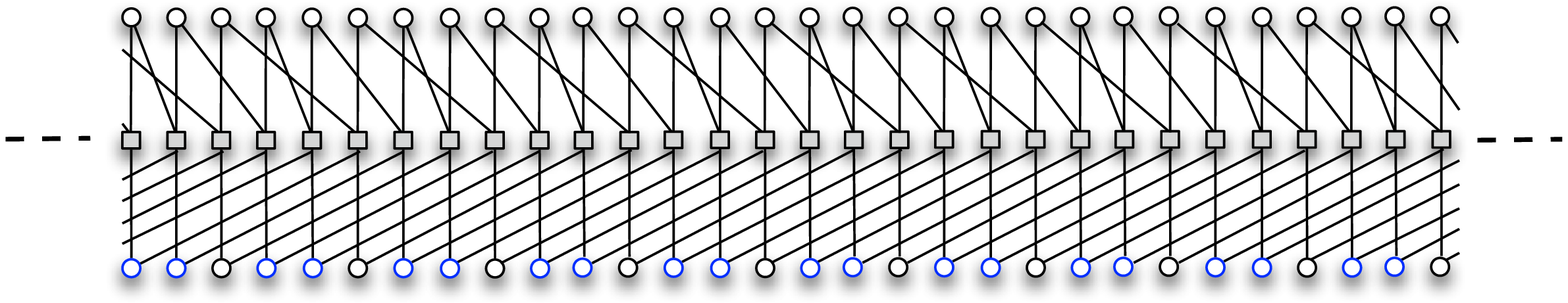}
\caption{Tanner graph of a (5,2,4)-regular LDPC convolutional code \self{of
 rate 3/4 with puncture bits. Blue nodes are punctured nodes.}}
\label{fig:punTg}
\end{figure}

\begin{figure}[h]
\centering
 \includegraphics[width=0.5\textwidth]{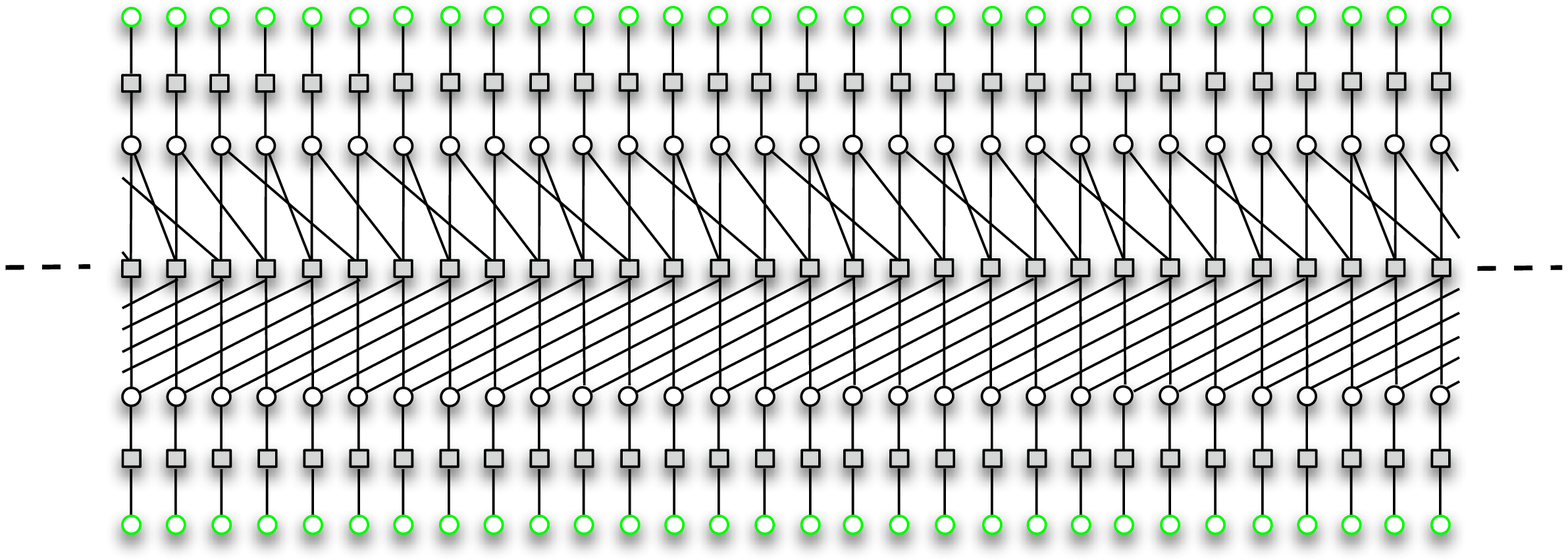}
\caption{Tanner graph of a (5,2,4)-regular LDPC convolutional code \self{of 
 rate 1/4 with multiplicative repetition. Green nodes are
 multiplicative repetition nodes.}}
\label{fig:repTg}
\end{figure}

\revB{Figures \ref{fig:punTg} and \ref{fig:repTg} describe the Tanner graph of a (5,2,4)-regular LDPC convolutional code used 
for decoding procedure with punctured symbols and multiplicatively repeated symbols, respectively. 
The coding rates are 3/4 and 1/4, respectively. }

\self{For the puncturing case in Fig. \ref{fig:punTg}, channel likelihoods of the puncturing nodes (blue
nodes) are initialized with uniform probability, then the iterative decoding process proceeds on the mother Tanner graph.
For the multiplicative repetition case in Fig. \ref{fig:repTg}, multiplicatively repeated nodes (green nodes) send each message just once
before the decoding process starts, then iterative decoding proceeds on the mother Tanner graph. 
In both cases, the decoder uses only the mother Tanner graph.
Hence, we do not need to change the decoder architecture for all rates.}

\begin{figure}[h]
\centering
\includegraphics[width=0.5\textwidth]{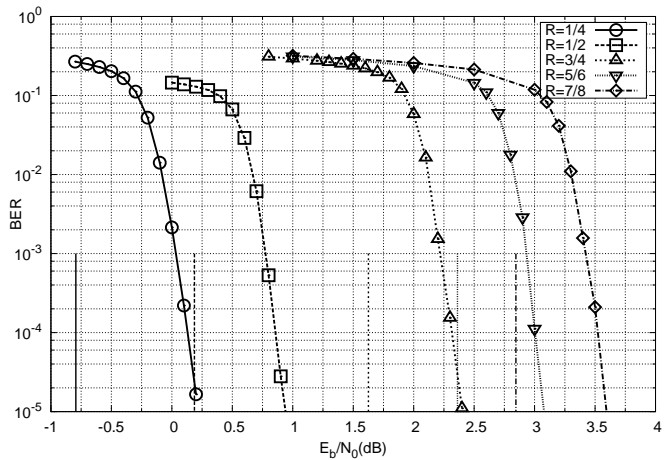}
\caption{Simulation results for rate-compatibility of non-binary LDPC convolutional
 codes over GF$(2^8)$ of rates 1/4, 1/2, 3/4, 5/6, and 7/8
 (marked curves). Corresponding Shannon limits to the rates are also
 described (vertical lines).}
\label{fig:rate_compati}
\end{figure}

Figure \ref{fig:rate_compati} shows the performance of rate-compatible non-binary LDPC convolutional codes. 
The mother code is a (52, 2, 4) LDPC convolutional code over GF$(2^8)$, which is evaluated in Section \ref{sec:sim}. 

The mother code of the rate 1/2, \self{multiplicatively} repeated code of the rate 1/4,
puncturing codes for rates = 3/4, 5/6, and 7/8 have bit error rate $10^{-4}$
around at $E_{b}/N_{0}$ = 0.9 dB, 0.05 dB, 2.2 dB, 3 dB, and 3.5 dB,
while the Shannon limits of the binary-input AWGN for rates 1/2, 1/4,
3/4, 5/6, and 7/8 are 0.187 dB, -0.794 dB, 1.626 dB, 2.362 dB, and 2.845
dB, respectively.
\self{As shown in} these curves, the proposed rate-compatible non-binary convolutional
code, although simple, can be constructed from low rates to high rates 
without large loss from the Shannon limits.


\begin{table*}[htb] 
\begin{center}
\caption{Puncturing patterns of R= 3/4, 5/6, and 7/8. \self{The} $\star$
 \self{represents the punctured}
 symbol and R=1/2 is a reference.}
\label{tb:punc_pat}
\begin{tabular}{|c|c|}
\hline
 R & sequence \\
\hline
 1/2 & $v_{t}^{(1)}, v_{t}^{(2)} , v_{t+1}^{(1)}, v_{t+1}^{(2)}, 
     v_{t+2}^{(1)}, v_{t+2}^{(2)}, \ldots$\\ \hline 
 3/4 & $v_{t}^{(1)}, \star, v_{t+1}^{(1)}, \star, v_{t+2}^{(1)}, 
     v_{t+2}^{(2)}, \ldots$ \\ \hline 
 5/6 & $v_{t}^{(1)}, \star, v_{t+1}^{(1)}, \star, v_{t+2}^{(1)}, \star,
     v_{t+3}^{(1)}, \star, v_{t+4}^{(1)}, v_{t+4}^{(2)}, \ldots$ \\ \hline 
 7/8 & $v_{t}^{(1)}, \star, v_{t+1}^{(1)}, \star, v_{t+2}^{(1)}, \star, 
     v_{t+3}^{(1)}, \star, v_{t+4}^{(1)}, \star, v_{t+5}^{(1)}, \star,
     v_{t+6}^{(1)}, v_{t+6}^{(2)}, \ldots$\\
\hline
\end{tabular}
\end{center}
\end{table*}

\section{Conclusions}
\label{sec:conclusions}
In this paper, we introduced terminated non-binary low-density 
parity-check (LDPC) convolutional codes and gave a construction
method of a syndrome former matrix.
Moreover we discussed the rate-compatibility of the
non-binary LDPC convolutional codes.
\revB{Simulation results showed that non-binary LDPC 
convolutional codes of rate 1/2 outperform binary
LDPC convolutional codes with smaller decoding latency.
Also the derived non-binary LDPC convolutional codes 
have good performance for rates from 1/4 to 7/8 
without large loss from the Shannon limits.}

However the non-binary LDPC block code outperforms the 
corresponding non-binary LDPC convolutional code for 
the same memory requirements. 
The density evolution results implied that
it is because the MAP threshold and the BP threshold are 
very close for $J=2$ regular non-binary LDPC codes. 
Since the BP threshold of (2,4) regular LDPC convolutional codes seems
to be slightly higher than that of the block code at the large field size, 
we believe the threshold saturation could be
observed with sufficiently large lengths.

\bibliographystyle{IEEEtran}
\bibliography{gotz-ref}

\end{document}